%% file: BDDgamma.tex
\documentclass[12pt]{article}
\usepackage{graphicx}
\def\beq{\begin{equation}}
\def\eeq{\end{equation}}
\def\bea{\begin{eqnarray}}
\def\eea{\end{eqnarray}}
\def\nn{\nonumber}
\def\sss{\scriptscriptstyle}
\def\roughly#1{\mathrel{\raise.3ex\hbox
{$#1$\kern-.75em\lower1ex\hbox{$\sim$}}}}

\def\bra#1{\left\langle #1\right|}
\def\ket#1{\left| #1\right\rangle}
\def\bd{B_d^0}

\def\bdbar{{\bar B}_d^0}
\def\btod{{\bar b} \to {\bar d}}
\def\btos{{\bar b} \to {\bar s}}
\def\ks{K_{\sss S}}

\def\Act{{\cal A}_{ct}}
\def\Aut{{\cal A}_{ut}}

\def\aR{a_{\sss R}}
\def\Actsig{{\cal A}_{ct,\sigma}}
\def\Autsig{{\cal A}_{ut,\sigma}}
\def\ActsigF{{\cal A}_{ct,\sigma}^F}
\def\AutsigF{{\cal A}_{ut,\sigma}^F}

\def\Acttwid{{\tilde{\cal A}}_{ct}}
\def\Auttwid{\tilde{{\cal A}}_{ut}}
\def\newprd#1#2#3{{ Phys.\ Rev.} {\bf D#1}, #3 (#2)}
\input babarsym
\input DDsymbols

\def\BtoDsDbar	{\ensuremath{B \to D^{(*)}_{\sss S} \Dbar^{(*)}}\xspace}

\begin{document}

\begin{flushright}  
CALT-68-2524 \\
UdeM-GPP-TH-04-127 \\
\end{flushright}

\begin{center}
\bigskip
{\Large \bf A Measurement of $\gamma$ from the Decays \vskip2truemm 
$\bd(t) \to D^{(*)+} D^{(*)-}$ and $\bd \to D_s^{(*)+} D^{(*)-}$} \\
\bigskip
\bigskip
{\large Justin Albert $^{a,}$\footnote{justin@hep.caltech.edu},
Alakabha Datta $^{b,}$\footnote{datta@physics.utoronto.ca} and David
London $^{c,}$\footnote{london@lps.umontreal.ca}}
\end{center}

\begin{flushleft}
~~~~~~~~~~~$a$: {\it Department of Physics, California Institute of
Technology,} \\
~~~~~~~~~~~~~~~{\it 1200 East California Blvd. Pasadena, CA 91125
USA}\\
~~~~~~~~~~~$b$: {\it Department of Physics, University of Toronto,}\\
~~~~~~~~~~~~~~~{\it 60 St.\ George Street, Toronto, ON, Canada M5S
1A7}\\
~~~~~~~~~~~$c$: {\it Laboratoire Ren\'e J.-A. L\'evesque, Universit\'e
de Montr\'eal,}\\
~~~~~~~~~~~~~~~{\it C.P. 6128, succ. centre-ville, Montr\'eal, QC,
Canada H3C 3J7}
\end{flushleft}

\begin{center} 
\bigskip (\today)
\vskip0.5cm
{\Large Abstract\\}
\vskip3truemm
\parbox[t]{\textwidth} {Recently, it was proposed to use measurements
of $\bd(t) \to D^{(*)+} D^{(*)-}$ and $\bd \to D_s^{(*)+} D^{(*)-}$
decays to measure the CP phase $\gamma$. In this paper, we present the
extraction of $\gamma$ using this method. We find that $\gamma$ is
favored to lie in one of the ranges $[19.4^{\circ}-80.6^{\circ}] (+
0^{\circ} \mbox{ or } 180^{\circ})$, $[120^{\circ}-147^{\circ}] (+
0^{\circ} \mbox{ or } 180^{\circ})$, or $[160^{\circ}-174^{\circ}] (+
0^{\circ} \mbox{ or } 180^{\circ})$ at 68\% confidence level (the $(+
0^{\circ} \mbox{ or } 180^{\circ})$ represents an additional ambiguity
for each range). These constraints come principally from the
vector-vector final states; the vector-pseudoscalar decays improve the
results only slightly. Although, with present data, the constraints
disappear for larger confidence levels, this study does demonstrate
the feasibility of the method. Strong constraints on $\gamma$ can be
obtained with more data.}
\end{center}

\thispagestyle{empty}
\newpage
\setcounter{page}{1}
% Decrease texheight (for preprint numbers) again
%\textheight 23.0 true cm
\baselineskip=14pt

Recently, two of us (AD, DL) proposed a method for extracting the CP
phase $\gamma$ from measurements of $\bd(t) \to D^{(*)+} D^{(*)-}$ and
$\bd \to D_s^{(*)+} D^{(*)-}$ decays \cite{BDD}. The technique is
quite straightforward. Consider the pseudoscalar-pseudoscalar (PP)
decay $\bd \to D^+ D^-$. The amplitude for this decay receives several
contributions, described by tree, exchange, $\btod$ penguin and
color-suppressed electroweak penguin diagrams \cite{su3}:
\bea
\label{AmpDDbar}
A^D & = & (T + E + P_c) \, V_{cb}^* V_{cd} + P_u \, V_{ub}^* V_{ud} +
(P_t + P_{\sss EW}^C) \, V_{tb}^* V_{td} \nn\\
& = & (T + E + P_c - P_t - P_{\sss EW}^C) \, V_{cb}^* V_{cd} + (P_u - P_t -
P_{\sss EW}^C) \, V_{ub}^* V_{ud} \nn\\
& \equiv & \Act\ e^{i \delta^{ct}} + \Aut\ e^{i \gamma} e^{i
\delta^{ut}} ~.
\eea
Here, $\Act\equiv |(T + E + P_c - P_t - P_{\sss EW}^C) V_{cb}^* V_{cd}|$,
$\Aut\equiv |(P_u - P_t - P_{\sss EW}^C) V_{ub}^* V_{ud}|$, $P_i$ is the
$\btod$ penguin amplitude with an internal $i$-quark, and we have
explicitly written out the strong phases $\delta^{ct}$ and
$\delta^{ut}$, as well as the weak phase $\gamma$. The second line is
obtained by using the unitarity of the CKM matrix, $V_{ub}^* V_{ud} +
V_{cb}^* V_{cd} + V_{tb}^* V_{td} = 0$, to eliminate the $V_{tb}^*
V_{td}$ term. The amplitude ${\bar A}^D$ for the decay $\bdbar \to D^+
D^-$ can be obtained from the above by changing the signs of the weak
phases.

There are three observables which can be obtained from a
time-dependent measurement of this decay: $B$ (the branching ratio),
$a_{dir}$ (the direct CP asymmetry) and $a_{indir}$ (the indirect CP
asymmetry). In terms of the above parameters, these can be written
\bea
\label{obsdefs}
B &\equiv & \frac{1}{2} \left( |A^D|^2 + |{\bar A}^D|^2 \right) =
\Act^2 + \Aut^2 + 2 \Act \Aut \cos\delta \cos\gamma ~, \nn \\
a_{dir} &\equiv & \frac{1}{2} \left( |A^D|^2 - |{\bar A}^D|^2 \right)
= - 2 \Act \Aut \sin\delta \sin\gamma ~, \\
a_{indir} &\equiv & {\rm Im}\left( e^{-2i \beta} {A^D}^* {\bar A}^D
\right) \nn\\
& = & -\Act^2 \sin 2\beta - 2 \Act \Aut \cos\delta \sin (2 \beta +
\gamma) - \Aut^2 \sin (2\beta + 2 \gamma)~, \nn
\eea
where ${\delta}\equiv {\delta}^{ut} - {\delta}^{ct}$. Here, $\beta$ is
the phase of $\bd$--$\bdbar$ mixing, which has been measured in the CP
asymmetry in $\bd(t)\to J/\psi \ks$ \cite{HFAG}. However, these
observables still depend on four unknown theoretical parameters: the
two magnitudes $\Act$ and $\Aut$, one relative strong phase
${\delta}$, and the weak phase $\gamma$. We therefore have three
observables, but four theoretical unknowns. Thus, in order to obtain
weak phase information, it is necessary to add some theoretical input
\cite{CKMambiguity}.

This input comes from the decay $\bd \to D_s^+ D^-$, which receives
tree, $\btos$ penguin and color-suppressed electroweak penguin
contributions \cite{su3}:
\bea
\label{AmpDsDbar}
A^{D_s} & = & (T' + P'_c) \, V_{cb}^* V_{cs} + P'_u \, V_{ub}^* V_{us}
+ (P'_t + P_{\sss EW}^{\prime C})\, V_{tb}^* V_{ts} \nn\\
& = & (T' + P'_c - P'_t - P_{\sss EW}^{\prime C}) \, V_{cb}^* V_{cs} +
(P'_u - P'_t - P_{\sss EW}^{\prime C}) \, V_{ub}^* V_{us} \nn\\
& \approx & (T' + P'_c - P'_t - P_{\sss EW}^{\prime C}) \, V_{cb}^* V_{cs}
\equiv \Act' e^{i \delta^{\prime ct}} ~.
\eea
(The primes on the amplitudes indicate a $\btos$ transition.)  Here,
the last line arises from the fact that $\left\vert {V_{ub}^* V_{us} /
V_{cb}^* V_{cs}} \right\vert \simeq 2\%$, so that the piece
proportional to $V_{ub}^* V_{us}$ is negligible. The measurement of
the total rate for $\bd \to D_s^+ D^-$ therefore yields $\Act'$.

We now make the SU(3) flavour assumption that
\beq
\Delta \equiv {\sin\theta_c \, \Act' \over \Act} = {\sin\theta_c |(T'
+ P'_c - P'_t - P_{\sss EW}^{\prime C}) V_{cb}^* V_{cs}| \over |(T + E +
P_c - P_t - P_{\sss EW}^C) V_{cb}^* V_{cd}|} = 1 ~,
\label{assumption1}
\eeq
where $\sin\theta_c$ is the Cabibbo angle. With this assumption, the
knowledge of $\Act'$ can be used to give us $\Act$. This in turn means
that the three observables in $\bd \to D^+ D^-$ depend on only three
theoretical unknowns: $\Aut$, $\delta$, and $\gamma$. We can therefore
now solve for $\gamma$ (up to discrete ambiguities).

The explicit solution for $\gamma$ is as follows. We introduce a
fourth (non-independent) observable, $\aR$:
\bea
\aR & \equiv & {\rm Re}\left( e^{-2i \beta} {A^D}^* {\bar A}^D
\right) = B^2 - a_{dir}^2 - a_{indir}^2 \nn\\
& = & \Act^2 \cos 2\beta + 2 \Act \Aut \cos\delta \cos (2 \beta +
\gamma) + \Aut^2 \cos (2\beta + 2 \gamma)~.
\eea
One can obtain $\aR$ from measurements of $B$, $a_{dir}$ and
$a_{indir}$, up to a sign ambiguity. One can then easily obtain
\beq
\Act^2 = { \aR \cos(2\beta + 2\gamma) - a_{indir} \sin(2\beta +
2\gamma) - B \over \cos 2\gamma - 1} ~.
\label{gammasolve}
\eeq
Given the knowledge of $2\beta$, the assumption in
Eq.~(\ref{assumption1}) therefore allows us to obtain $\gamma$.

The leading-order theoretical error in this technique is given simply
by the SU(3)-breaking ratio of decay constants $f_{D_s}/f_D$
\cite{BDD}. (There are other errors, such as the neglect of the $E$
amplitude in $\Act$, but these are expected to be smaller.) This ratio
has been computed quite precisely on the lattice: $f_{D_s}/f_D = 1.20
\pm 0.06 \pm 0.06$ \cite{lattice}. With this value, the theoretical
error in this method is rather small, so that $\gamma$ can be
extracted from measurements of $\bd(t) \to D^+ D^-$ and $\bd \to D_s^+
D^-$.

Unfortunately, at present data on $\bd(t) \to D^+ D^-$ is unavailable,
so we cannot apply the above method to this decay. However, the
vector-vector (VV) decays $\bd(t) \to D^{*+} D^{*-}$ and $\bd \to
D_s^{*+} D^{*-}$ {\it have} been measured. The method can be applied
in a similar way to these decays, but with some additional complexity
as described below. The main theoretical error is given by
$f_{D^*_s}/f_{D^*}$.

A modification of this method can be used when the final state is not
self-conjugate, as is the case for vector-pseudoscalar (VP) final
states \cite{BKKbar}. Consider the decay $\bd \to D^{*+} D^-$.
Following Eq.~(\ref{AmpDDbar}), its amplitude can be written
\beq
A(\bd\to D^{*+} D^-) = \Act\ e^{i \delta^{ct}} + \Aut\ e^{i \gamma}
e^{i \delta^{ut}} ~.
\eeq
(Although we use the same symbols, the amplitudes and strong phases
are not the same as those for $\bd \to D^+ D^-$.) Now consider the
decay of a $\bdbar$ meson to the same final state, $\bdbar \to D^{*+}
D^-$. The amplitude for this decay is not simply related to that for
$\bd \to D^{*+} D^-$ since the hadronization is different: in $\bd \to
D^{*+} D^-$, the spectator quark is part of the $D^-$, while in
$\bdbar \to D^{*+} D^-$ it is contained in the $D^{*+}$. We therefore
write
\beq
A(\bdbar\to D^{*+} D^-) = \Acttwid\ e^{i {\tilde\delta}^{ct}} +
\Auttwid\ e^{-i \gamma} e^{i {\tilde\delta}^{ut}} ~.
\eeq

The measurement of $\bd(t) \to D^{*+} D^-$ still yields three
observables, $B$, $a_{dir}$ and $a_{indir}$, but now they take more
complicated forms. The three observables for the $D^{*+} D^-$ final
state are:
\bea
\label{VP1}
B & = & \frac12 \left[ \Aut^2 + \Act^2 + 2 \Aut \Act \cos(\gamma +
\delta) + {\Auttwid}^2 + {\Acttwid}^2 + 2 \Auttwid \Acttwid
\cos(\gamma - {\tilde\delta}) \right] ~, \nn\\
a_{dir} & = & \frac12 \left[ \Aut^2 + \Act^2 + 2 \Aut \Act \cos(\gamma
+ \delta) - {\Auttwid}^2 - {\Acttwid}^2 - 2 \Auttwid \Acttwid
\cos(\gamma - {\tilde\delta}) \right] ~, \nn\\
a_{indir} & = & \Aut \Auttwid \sin(-2 \beta - 2 \gamma - \delta +
{\tilde\delta} - \Delta) + \Aut \Acttwid \sin( -2 \beta - \gamma -
\delta - \Delta) \nn\\
& & \hskip1truecm + \Act \Auttwid \sin( -2 \beta - \gamma +
{\tilde\delta} - \Delta) + \Act \Acttwid \sin( -2 \beta - \Delta) ~,
\eea
where ${\delta}\equiv {\delta}^{ut} - {\delta}^{ct}$, ${\tilde\delta}
\equiv {\tilde \delta}^{ut} - {\tilde \delta}^{ct}$, and $\Delta
\equiv \delta_{ct} - {\tilde\delta}_{ct}$.

For the $ D^+ D^{*-}$ final state, the observables are
\bea
\label{VP2}
{\tilde B} & = & \frac12 \left[ {\Auttwid}^2 + {\Acttwid}^2 + 2 \Auttwid
\Acttwid \cos(\gamma + {\tilde\delta}) + \Aut^2 + \Act^2 + 2 \Aut \Act
\cos(\gamma - \delta) \right] ~, \nn\\
{\tilde a}_{dir} & = & \frac12 \left[ {\Auttwid}^2 + {\Acttwid}^2 + 2 \Auttwid
\Acttwid \cos(\gamma + {\tilde\delta}) - \Aut^2 - \Act^2 - 2 \Aut \Act
\cos(\gamma - \delta) \right] ~, \nn\\
{\tilde a}_{indir} & = & \Aut \Auttwid \sin(-2 \beta - 2 \gamma + \delta -
{\tilde\delta} + \Delta) + \Aut \Acttwid \sin( -2 \beta - \gamma +
\delta + \Delta) \nn\\
& & \hskip1truecm + \Act \Auttwid \sin( -2 \beta - \gamma -
{\tilde\delta} + \Delta) + \Act \Acttwid \sin( -2 \beta + \Delta) ~.
\eea

Considering that $\beta$ has been experimentally determined, the 6
observables are written in terms of 8 theoretical unknowns: $\Aut$,
$\Act$, $\Auttwid$, $\Acttwid$, $\gamma$, $\delta$, ${\tilde\delta}$
and $\Delta$. We therefore need 2 assumptions to extract
information. These come from using the decays $\bd\to D_s^{*+} D^-$
and $\bd\to D_s^+ D^{*-}$. The measurements of the branching ratios
for these decays allow us to extract ${\cal A}'_{ct}$ and
${\tilde{\cal A}}'_{ct}$, respectively. With motivation similar to
that for the PP mode above, we assume that
\beq
{\sin\theta_c \, {\cal A}'_{ct} \over \Act} = 1 ~,~~
{\sin\theta_c \, {\tilde{\cal A}}'_{ct} \over \Acttwid} = 1 ~.
\eeq
With these assumptions, the measurements of the branching ratios for
$\bd\to D_s^{*+} D^-$ and $\bd\to D_s^+ D^{*-}$ give us $\Act$ and
$\Acttwid$. We now have 6 observables and 6 theoretical unknowns, thus
we can solve for $\gamma$ (up to discrete ambiguities). The main
theoretical error is the deviation from unity of $f_{D_s^*}/f_{D^*}$
and $f_{D_s}/f_D$ in the first and second assumption above,
respectively.

In this Letter we extract $\gamma$ via this method, using BaBar and
Belle data on the VV and VP modes \cite{VVdata1}--\cite{VVdata7}. We
determine the value of $\gamma$ to be in one of the ranges
$[19.4^{\circ}-80.6^{\circ}] (+ 0^{\circ} \mbox{ or } 180^{\circ})$,
$[120^{\circ}-147^{\circ}] (+ 0^{\circ} \mbox{ or } 180^{\circ})$, or
$[160^{\circ}-174^{\circ}] (+ 0^{\circ} \mbox{ or } 180^{\circ})$ at
68\% confidence level (C.L.), where the $(+ 0^{\circ} \mbox{ or }
180^{\circ})$ represents an additional ambiguity for each range. The
first of these ranges, $[19.4^{\circ}-80.6^{\circ}]$, is the range
that is favored by other, external information on the Unitarity
Triangle (assuming the standard model)
\cite{CKMfitter,UTfitter,UnFit}. As we will see, the constraints on
$\gamma$ come principally from the data on VV modes; at present, the
VP decays add only a small amount. Note also that if we consider a
larger C.L.  (e.g.\ 90\%), the constraints on $\gamma$
disappear. Thus, even though we obtain limits on $\gamma$ at 68\%
C.L., our main purpose here is to demonstate the feasibility of the
method. With more data, the constraints on $\gamma$ will be
correspondingly stronger. Eventually this method can be used to obtain
a precision determination of $\gamma$.

We begin with the analysis of vector-vector decays. In order to
extract $\gamma$, however, we must make additional assumptions. VV
final states come in three transversity states, $0$, $\|$, and
$\perp$. The amplitudes $0$ and $\|$ are CP-even, while $\perp$ is
CP-odd. The data shows that the $D^* {\bar D}^*$ final state is almost
entirely CP-even, i.e.\ the $\perp$ amplitude is negligible
\cite{VVdata5}. Unfortunately, at present experiments cannot
distinguish between the $0$ and $\|$ amplitudes. This has the
following effect. For a single transversity $\sigma$, the three
observables in $\bd(t) \to D^{*+} D^{*-}$ can be written similarly to
Eq.~(\ref{obsdefs}):
\bea
\label{D*D*obs}
B_{\sigma} & = & \Actsig^2 + \Autsig^2 + 2 \, \Actsig \,
\Autsig \cos\delta_{\sigma} \cos\gamma ~, \nn \\
a_{dir}^\sigma & = & - 2 \, \Actsig \, \Autsig
\sin\delta_{\sigma} \sin\gamma ~, \\
a_{indir}^\sigma & =  & -\Actsig^2 \sin 2\beta - 2 \, \Actsig \,
\Autsig \cos\delta_{\sigma} \sin (2 \beta + \gamma) - \Autsig^2 \sin
(2\beta + 2 \gamma)~, \nn
\eea
where ${\delta_{\sigma}\equiv {\delta}^{ut}_{\sigma} -
{\delta}^{ct}_{\sigma}}$. However, for all three observables, what is
measured is the {\it sum} of the helicities $0$ and $\|$:
\bea
\label{obsVVdefs}
B & \!=\! & \sum_{\sigma = 0,\|} \left[ \Actsig^2
+ \Autsig^2 + 2 \, \Actsig \, \Autsig \cos\delta_{\sigma} \cos\gamma
\right], \nn \\
a_{dir } & \!=\! & \sum_{\sigma = 0,\|} \left[ -2 \Actsig \, \Autsig
  \sin\delta_{\sigma} \sin\gamma \right] ~, \\
a_{indir} & \!=\! & \sum_{\sigma = 0,\|} \left[ -\Actsig^2 \sin 2\beta
- 2 \, \Actsig \, \Autsig \cos\delta_{\sigma} \sin (2 \beta + \gamma)
- \Autsig^2 \sin (2\beta + 2 \gamma) \right]. \nn
\eea
Because the parameters $\Act$, $\Aut$ and ${\delta}$ are different for
the two helicities $0$ and $\|$, there are too many theoretical
unknowns to apply the method.

We would like to cast the expressions for the observables in
Eq.~(\ref{obsVVdefs}) in the same form as those in
Eq.~(\ref{obsdefs}). In order to do this, we must relate the
parameters for the $0$ and $\|$ helicities. Specifically, we assume
that the strong phases are equal: $\delta_0 = \delta_\| \equiv
\delta$. We also assume that the amplitudes are proportional to one
another (with the same proportionality constant, $c$): ${\cal
A}_{ct,0} = c {\cal A}_{ct,\|}$, $ {\cal A}_{ut,0} = c {\cal
A}_{ut,\|} $. That is, our assumptions are:
\beq
{ {\cal A}_{ct,0} \over {\cal A}_{ct,\|} } = { {\cal A}_{ut,0} \over
{\cal A}_{ut,\|} } ~,~~~ \delta_0 = \delta_\| ~.
\label{assumps}
\eeq
With these definitions, the observables in Eq.~(\ref{obsVVdefs}) take
exactly the same form as Eq.~(\ref{obsdefs}), and the method for
extracting $\gamma$ can be applied.

In fact, these assumptions are theoretically reasonable. The
amplititudes for $\bd \to D^* \bar{D}^*$ and $\bdbar \to \bar{D}^*
{D}^*$ are given by
\beq
A_{VV, \sigma} = \Actsig e^{i {\delta}^{ct}_{\sigma}} + \Autsig
e^{i {\delta}^{ut}_{\sigma}} e^{ i \gamma} ~~,~~~~
{\bar A}_{VV, \sigma} = \Actsig e^{i {\delta}^{ct}_{\sigma}} +
\Autsig e^{i {\delta}^{ut}_{\sigma}} e^{ -i \gamma} ~,
\label{Dstaramp}
\eeq
with $\sigma =0, \|, \perp$ being the three transversity states. We
now define the ratios of amplitudes:
\beq
k_{ct} \equiv { {\cal A}_{ct,0} \over {\cal A}_{ct,\|} } e^{i
({\delta}^{ct}_0 - {\delta}^{ct}_\|)}
~~,~~~~ k_{ut} \equiv { {\cal A}_{ut,0} \over {\cal A}_{ut,\|} } e^{i
({\delta}^{ut}_0 - {\delta}^{ut}_\|)} ~.
\eeq
Our assumptions are equivalent to $k_{ct} = k_{ut}$, i.e.\ both
amplitudes and phases are equal. Below, we investigate the extent to
which these relations hold true.

We note that, in general, we can write any amplitude in terms of
factorizable and nonfactorizable pieces:
\beq
\Actsig = \ActsigF e^{i {\delta}^{ct,F}_{\sigma}} \left [ 1+ r_\sigma
e^{ i \rho_\sigma} \right] ~~,~~~~
\Autsig = \AutsigF e^{i {\delta}^{ut,F}_{\sigma}} \left [ 1+ s_\sigma
e^{ i \lambda_\sigma} \right] ~,
\label{fac-nonfac}
\eeq
where we denote the factorizable contributions by an index `$F$'. The
quantities $r_\sigma$, $\rho_\sigma$, $s_\sigma$ and $\lambda_\sigma$
parametrize the ratios of nonfactorizable and factorizable amplitudes.

Consider first only the factorizable contributions. In this case, the
ratios of factorizable amplitudes are
\beq
k_{ct}^F \equiv { {\cal A}_{ct,0}^F \over {\cal A}_{ct,\|}^F } e^{i
({\delta}^{ct,F}_0 - {\delta}^{ct,F}_\|)}
~~,~~~~ k_{ut}^F \equiv { {\cal A}_{ut,0}^F \over {\cal A}_{ut,\|}^F }
e^{i ({\delta}^{ut,F}_0 - {\delta}^{ut,F}_\|)} ~.
\eeq
The factorizable amplitude for the decay $\bdbar \to \bar{D}^* D^*$ is
given by \cite{luorosner}
\beq
A[\bd \to D^{*} \bar{D}^*]^\sigma = \frac{G_F}{\sqrt{2}} X
P_{D^*}^{\sigma},
\label{ampfac1}
\eeq
where $X=X_1 +X_2$, and
\bea
\label{ampfac2}
X_1 & = & V_{cb} V_{cd}^* \left[a_2 +a_4^t+a_{10}^t -a_4^c-a_{10}^c
 \right] ~, \nn\\
X_2 & = & V_{ub} V_{ud}^* \left[a_4^t+a_{10}^t -a_4^u-a_{10}^u \right]
~, \nn\\
P_{D^*}^\sigma & = & \left[m_{D^*} f_{D^*} \varepsilon^{*\mu}_{D^*}
\bra{D^*} \bar{c} \gamma_{\mu} (1 - \gamma_5) b
\ket{\bdbar}\right]^{\sigma} ~.
\eea
In the above, $a_j=c_j+ c_{j-1}/N_c$, where the $c_j$ are Wilson
coefficients. From Eqs.~(\ref{ampfac1}) and (\ref{ampfac2}), we can
read off the individual factorizable amplitudes:
\beq
\ActsigF e^{i {\delta}^{ct,F}_{\sigma}} = \frac{G_F}{\sqrt{2}} X_1
P_{D^*}^{\sigma} ~~,~~~~
\AutsigF e^{i {\delta}^{ut,F}_{\sigma}} = \frac{G_F}{\sqrt{2}} X_2
P_{D^*}^{\sigma} ~.
\label{fac_amp}
\eeq
However, note that the strong phases come from $a_{4,10}^{u,c}$, and
appear only in the factors $X_{1,2}$ in Eq.~(\ref{ampfac2}). These
factors are independent of transversity. That is, the relative strong
phases between the factorizable $ct$ and $ut$ amplitudes are
independent of the polarization state, leading to ${\rm Arg}
(k^F_{ct}) = {\rm Arg} (k_{ut}^F)$. Furthermore, the expressions in
Eq.~(\ref{fac_amp}) above lead to
\beq
\left\vert k^F_{ct} \right\vert = \left\vert k^F_{ut} \right\vert =
\frac{P_{D^*}^{0}}{ P_{D^*}^{\|}} ~.
\eeq
Thus, we have $k^F_{ct} = k_{ut}^F$, i.e.\ the factorizable
contributions satisfy our assumptions.

We now consider the nonfactorizable contributions. If these pieces are
independent of polarization (at least for the $\sigma = 0, \|$
states), then in Eq.~(\ref{fac-nonfac}) we will have
\beq
r_0 = r_\| \equiv r ~~,~~ s_0 = s_\| \equiv s ~~,~~ \rho_0 = \rho_\|
\equiv \rho ~~,~~ \lambda_0 = \lambda_\| \equiv \lambda ~.
\label{nonfac_assumption}
\eeq
This leads to $k_{ct} = k_{ut}$, so that our assumptions will be
satisfied. Our assumptions are therefore invalid only to the extent
that the nonfactorizable pieces are transversity-dependent. 

In the heavy-quark limit with $m_{b,c} \to \infty$, there is only one
universal form factor resulting from the spin symmetry of the
theory. This implies that the (factorizable) $P_{D^*}^{\sigma}$'s in
Eq.~(\ref{fac_amp}) are proportional for different polarization
states. In other words, the various transversity amplitudes are
related to one another. It is likely that these relations remain true
in the presence of nonfactorizable corrections. This then implies
Eq.~(\ref{nonfac_assumption}). We therefore expect deviations from
Eq.~(\ref{nonfac_assumption}) to be suppressed by $O(1/m_{c,b})$. In
all, the net correction to our assumptions is $O(1/m_{c,b})$ times the
ratio of nonfactorizable and factorizable effects. We expect this to
be small, so that the assumptions in Eq.~(\ref{assumps}) are
justified.

Finally, we note that our assumptions can be tested. In the presence
of nonfactorizable effects of the form in
Eq.~(\ref{nonfac_assumption}), Eq.~(\ref{ampfac1}) can be rewritten as
\beq
A[\bd \to D^{*} \bar{D}^*]^\sigma = \frac{G_F}{\sqrt{2}} X_T P_{D^*}^{\sigma},
\eeq
where $X_T=X_{1T} +X_{2T}$, and
\bea
\label{amptotal}
X_{1T} & = & V_{cb} V_{cd}^* \left[a_2 +a_4^t+a_{10}^t -a_4^c-a_{10}^c
\right] \left[ 1+r e^{ i \rho} \right] ~, \nn\\
X_2 & = & V_{ub} V_{ud}^* \left[a_4^t+a_{10}^t -a_4^u-a_{10}^u \right]
\left[ 1+s e^{ i \lambda} \right] ~, \nn\\
P_{D^*}^\sigma & = & \left[m_{D^*} f_{D^*} \varepsilon^{*\mu}_{D^*}
\bra{D^*} \bar{c} \gamma_{\mu} (1 - \gamma_5) b
\ket{\bdbar}\right]^{\sigma} ~.
\eea
As $X_T$ is common to both $\sigma= 0, \|$ states we find the relative
phase between these transversity amplitudes is 0 or $\pi$. This
prediction can be checked through an angular analysis of $B \to D^*
{\bar D}^*$ decays. Note also that the assumptions in
Eq.~(\ref{assumps}) are not in fact required in the original method of
Ref.~\cite{BDD}. Eventually it will be possible to experimentally
separate out the $0$ and the $\|$ components, making such assumptions
unnecessary.

With the assumptions of Eq.~(\ref{assumps}), we can determine the
value of $\gamma$ from the VV decays, using the method above. We use
the measurements of the \BtoDDbar\ and \BtoDsDbar\ branching
fractions, measurements of the \BtoDDbar\ CP asymmetries, and the
world-average values of $\sin 2\beta$ ($0.736 \pm 0.049$)~\cite{HFAG}
and $\sin^2 \theta_c$ ($0.0482 \pm 0.0010$)~\cite{PDG2004}. We take
$2\beta$ to lie in the first quadrant. 

\begin{figure}[!t]
\begin{center}
\includegraphics[width=0.55\textwidth]{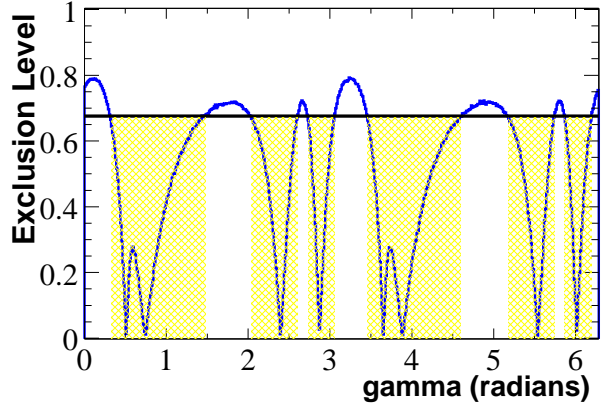}
\caption{ The measured exclusion level, as a function of $\gamma$,
from a fit to the vector-vector modes ($D^{*+}D^{*-}$ and
$D_s^{*+}D^{*-}$).  The exclusion level is defined in the text.  From
the fit, $\gamma$ is favored to lie in one of the ranges $[0.31-1.50]
(+ 0 \mbox{ or } \pi)$, $[2.02-2.62] (+ 0 \mbox{ or } \pi)$, or
$[2.72-3.05] (+ 0 \mbox{ or } \pi)$ radians at 68\% confidence level.}
\label{fig:chi2}
\end{center}
\end{figure}

We use a toy Monte Carlo (MC) method to determine the confidence
intervals for $\gamma$. We consider 500 values for $\gamma$, evenly
spaced between 0 and $2\pi$.  For each value of $\gamma$ considered,
we generate 25000 toy MC experiments, with inputs that span the range
of the experimental errors of each quantity. For each experiment, we
generate random values of each of the experimental inputs according to
Gaussian distributions, with means and sigmas according to the
measured central value and total errors on each experimental quantity.
We make the assumption that the ratio $f_{D_s^*}/f_{D^*}$ is equal to
$f_{D_s}/f_D = 1.20 \pm 0.06 \pm 0.06$ \cite{lattice}. An additional
theory error of 10\% is included to take into account the assumptions
of Eq.~(\ref{assumps}), as well as smaller errors such as the neglect
of the exchange diagram, subdominant SU(3)-breaking terms, etc. We
then calculate the resulting values of $\Act$, $a_{\rm dir}$, $a_{\rm
indir}$, and $B$, given the generated random values (based on the
experimental values). Inputting the quantities $a_{\rm dir}$, $a_{\rm
indir}$, and $B$, along with $\beta$ and the value of $\gamma$ that is
being considered, into Eq.~(\ref{gammasolve}), we obtain a residual
value for each experiment, equal to the difference of the left- and
right-hand sides of the equation. One thus obtains an ensemble of
residual values from the 25000 experiments. A likelihood, as a
function of $\gamma$, can be obtained from $\chi^2 (\gamma)$, where
$\chi^2 \equiv (\mu/\sigma)^2$, in which $\mu$ is the mean of the
above ensemble of residual values and $\sigma$ is the usual square
root of the variance.  The value of $\chi(\gamma)$ is then considered
to represent a likelihood which is equal to that of a value $\chi$
standard devations of a Gaussian distribution from the most likely
value(s) of $\gamma$.  We define the ``exclusion level,'' as a
function of the value of $\gamma$, as follows: the value of $\gamma$
is excluded from a range at a given C.L. if the exclusion level in
that range of $\gamma$ values is greater than the given C.L.

\begin{figure}[!t]
\begin{center}
\includegraphics[width=0.6\textwidth]{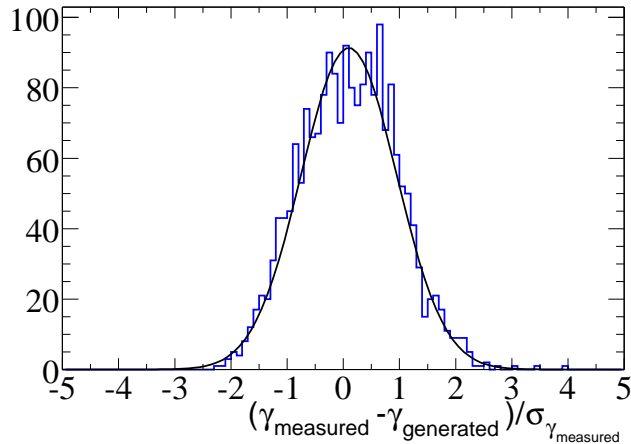}
\caption{A ``pull distribution'' check on the measured confidence
levels for the $\gamma$ fit.  Values of $\gamma$, as well as the other
theoretical parameters $\mathcal{A}_{ct}$, $\mathcal{A}_{ut}$,
$\delta$, and $\beta$, are generated and used to produce values of the
experimental inputs to the fit. The fit is then performed, and
((measured value of $\gamma$) $-$ (generated value of
$\gamma$))/(measured uncertainty on $\gamma$) is plotted. The result
is consistent with a Gaussian distribution with $\sigma = 1$, implying
that uncertainty on the measured value of $\gamma$ is accurately
described by the confidence distribution.}
\label{fig:pulldist}
\end{center}
\end{figure}

Fig.~\ref{fig:chi2} shows the resulting measured confidence as a
function of $\gamma$. We see that $\gamma$ is favored to lie in one of
the ranges $[0.31-1.50] (+ 0 \mbox{ or } \pi)$, $[2.02-2.62] (+ 0
\mbox{ or } \pi)$, or $[2.72-3.05] (+ 0 \mbox{ or } \pi)$ radians at
68\% C.L. This corresponds to $[18.0^{\circ}-85.7^{\circ}] (+
0^{\circ} \mbox{ or }$ $180^{\circ})$, $[116^{\circ}-150^{\circ}] (+
0^{\circ} \mbox{ or } 180^{\circ})$, or $[156^{\circ}-175^{\circ}] (+
0^{\circ} \mbox{ or } 180^{\circ})$.  Fig.~\ref{fig:pulldist} shows a
check on the confidence distribution to ensure that it accurately
describes the level of uncertainty on the measured value of $\gamma$.

We now turn to the VP decays $\bd(t) \to D^{*\pm} D^{\mp}$, $\bd \to
D_s^{*+} D^{-}$, and $\bd \to D_s^{+} D^{*-}$. The advantage of the VP
method is that no additional assumptions of the type described in
Eq.~(\ref{assumps}) are needed. The disadvantage is that, as we will
see, the data are such that no information on the most likely regions
of $\gamma$ can be obtained from the VP modes.

In order to implement the VP method, we proceed as follows. We first
use the expressions for $B$, $a_{dir}$, ${\tilde B}$ and ${\tilde
a}_{dir}$ in Eqs.~(\ref{VP1}) and (\ref{VP2}) to analytically solve
for the theoretical unknowns $\Aut$, $\Auttwid$, $\delta$, and
$\delta'$, up to a four-fold ambiguity. We then have the two remaining
equations for $a_{indir}$ and ${\tilde a}_{indir}$ and two theoretical
unknowns, $\gamma$ and $\Delta$. We now consider 200 values for each
of $\gamma$ and $\Delta$, each evenly spaced between 0 and $2\pi$.
For each of the $200 \times 200$ possible combinations of values of
$\gamma$ and $\Delta$, we generate 5000 toy MC experiments, with
inputs that span the range of the experimental errors of each
quantity.

\begin{figure}[!t]
\begin{center}
\includegraphics[width=0.55\textwidth]{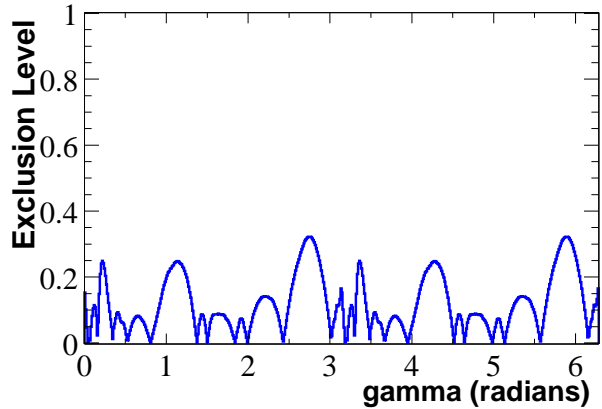}
\caption{ The measured exclusion level, as a function of $\gamma$,
from a fit to the vector-pseudoscalar modes $D^{*\pm} D^{\mp}$,
$D_s^{*+} D^{-}$, and $D_s^{+} D^{*-}$. Unlike the vector-vector
modes, with present data we do not obtain useful information on the
likely regions of $\gamma$ from these modes alone.  }
\label{fig:chi2_dstd}
\end{center}
\end{figure}

Similar to the toy Monte Carlo $\gamma$ determination for the VV mode,
we generate random values of each of the experimental inputs according
to Gaussian distributions, with means and sigmas according to the
measured central value and total errors on each experimental quantity.
Making the assumption that $f_{D_s^*}/f_{D^*}$ is equal to
$f_{D_s}/f_D = 1.20 \pm 0.06 \pm 0.06$ \cite{lattice}, we again obtain
a confidence distribution as a function of $\gamma$. The result is
shown in Fig.~\ref{fig:chi2_dstd}. As can been seen in this figure,
present data on VP decays alone do not lead to useful constraints on
$\gamma$.

\begin{figure}[!t]
\begin{center}
\includegraphics[width=0.55\textwidth]{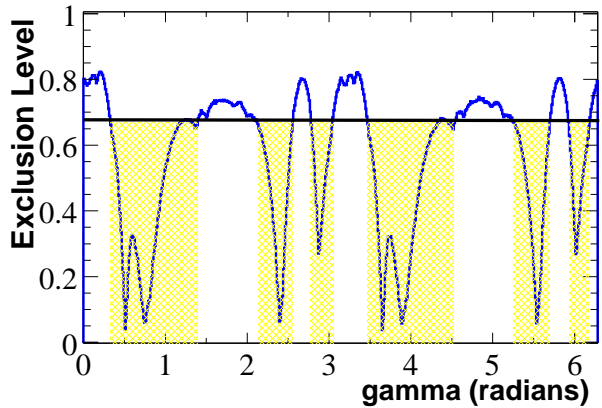}
\caption{ The measured exclusion level, as a function of $\gamma$,
from the combined information from vector-vector and
vector-pseudoscalar modes.  The combined information implies that
$\gamma$ is favored to lie in one of the ranges $[0.34-1.41] (+ 0
\mbox{ or } \pi)$, $[2.09-2.57] (+ 0 \mbox{ or } \pi)$, or
$[2.79-3.04] (+ 0 \mbox{ or } \pi)$ radians at 68\% confidence level.}
\label{fig:chi2combined}
\end{center}
\end{figure}

Finally, we can combine information from the VV and VP modes. The
result is shown in Fig.~\ref{fig:chi2combined}. From the combined fit,
we see that $\gamma$ is favored to lie in one of the ranges
$[0.34-1.41] (+ 0 \mbox{ or } \pi)$, $[2.09-2.57] (+ 0 \mbox{ or }
\pi)$, or $[2.79-3.04] (+ 0 \mbox{ or } \pi)$ radians at 68\%
confidence level. This corresponds to $[19.4^{\circ}-80.6^{\circ}] (+
0^{\circ} \mbox{ or } 180^{\circ})$, $[120^{\circ}-147^{\circ}] (+
0^{\circ} \mbox{ or } 180^{\circ})$, or $[160^{\circ}-174^{\circ}] (+
0^{\circ} \mbox{ or } 180^{\circ})$. Comparing Figs.~\ref{fig:chi2}
and \ref{fig:chi2combined}, we see that the favored ranges of $\gamma$
are slightly more constrained with the VP data. Thus, although the VP
data does not by itself constrain $\gamma$, its inclusion in the
combined fit does have an effect.

To summarize, we have presented the extraction of $\gamma$ using
measurements of $\bd(t) \to D^{(*)+} D^{(*)-}$ and $\bd \to D_s^{(*)+}
D^{(*)-}$ decays \cite{BDD}. We find that $\gamma$ is favored to lie
in one of the ranges $[19.4^{\circ}-80.6^{\circ}] (+ 0^{\circ} \mbox{
or } 180^{\circ})$, $[120^{\circ}-147^{\circ}] (+ 0^{\circ} \mbox{ or
} 180^{\circ})$, or $[160^{\circ}-174^{\circ}] (+ 0^{\circ} \mbox{ or
} 180^{\circ})$ at 68\% confidence level (the $(+ 0^{\circ} \mbox{ or
} 180^{\circ})$ represents an additional ambiguity for each range).
The first of these ranges is that favored by fits to the Unitarity
Triangle, assuming the standard model. The ranges come principally
from data on vector-vector decays, although the vector-pseudoscalar
decays do improve the constraints slightly. Note that, if we consider
a larger confidence level, there are no constraints on $\gamma$ at
present. However, this study demonstrates the feasibility of the
method -- with more data, we will be able to obtain strong constraints
on $\gamma$.

\bigskip
\noindent
{\bf Acknowledgements}:
%\bigskip
We thank Andreas Kronfeld for helpful communications regarding the
lattice values of $f_{D_s}/f_D$ and $f_{D_s^*}/f_{D^*}$. J.A. is
partially supported by DOE contract DE-FG01-04ER04-02. The work of
A.D. and D.L. is financially supported by NSERC of Canada.

%%%%%%%%%%%%%%%%%%%%% REFERENCES %%%%%%%%%%%%%%%%%%%%%%%%%%%%%%%%

\end{document}

%% file: DDsymbols.tex
\def\Dstarp     {\ensuremath{D^{*+}}\xspace}
\def\Dstarm     {\ensuremath{D^{*-}}\xspace}

\def\Dp         {\ensuremath{D^+}\xspace}
\def\Dm         {\ensuremath{D^-}\xspace}

\def\BtoDDbar	{\ensuremath{B \to D^{(*)} \Dbar^{(*)}}\xspace}

\def\BtoDsDbar	{\ensuremath{B \to D^{(*)}_S \Dbar^{(*)}}\xspace}

%% file: BDDgamma.bbl
\begin{thebibliography}{99}
\bibitem{BKKbar} A similar method is described in A. Datta and
D. London, Phys.\ Lett.\ B {\bf 533}, 65 (2002). However, there is a
mistake in that method, due to the fact that the hadronization is
different for the decays $\bd(t) \to K^* {\bar K}$ and $\bd(t) \to K
{\bar K}^*$.
\end{thebibliography}
